# Superconductivity in Mg/MgO interface

N.S. Sidorov, A.V. Palnichenko, O.M. Vyaselev

*Institute of Solid State Physics, Russian Academy of Sciences, Chernogolovka, Moscow District, 142432 Russia*

**Abstract**

Metastable superconductivity at ≈ 50 K in the interfaces formed by metallic and oxidized magnesium (MgO) has been observed by ac magnetic susceptibility and electrical resistance measurements. The superconducting interfaces have been produced by the surface oxidation of metallic magnesium under special conditions.

## 1. Introduction

Interfacial superconductivity, which can occur in a boundary layer between two different materials, even insulators [1], is a consequence of a large change of physical properties of these materials in the layer compared to their properties in the bulk [2]. The challenge to understand, predict, and tailor the superconducting properties of the interfaces is enormous [3, 4]. For example, interface superconductivity formed by different oxides is considered a promising opportunity to unravel the mystery of high-temperature superconductivity (HTS) [5] and increase the temperature of superconducting transition $T_c$ [6].

Despite the novel technical advances that enabled breakthrough experiments and led to a discovery of the interfacial superconductivity at the junction of two initially non-superconducting materials, there are only few examples of the interfacial superconducting layer with $T_c$ higher than in the bulk optimally doped samples [4].

The origin of the interfacial superconductivity is not completely understood yet. Moreover, the complexity of the used oxides complicates the issue [1–4, 6]. Therefore, in order to reveal essential factors responsible for the interface superconductivity, development of high-temperature superconducting interfaces between simplest substances would illuminate the HTS study.

We have reported previously that thermal treatment of magnesium diboride ($MgB_2$) samples with alkali metals results in increasing of the superconducting transition temperature of the samples up to 45-58 K [7]. It has been found later that it is the presence of magnesium oxide $MgO$ in the sample that is crucial for the enhancement of $T_c$; the same treatment of pure $MgB_2$ does not change the $T_c \approx 39$ K of the sample, while $T_c$ of the $MgB_2$ samples highly contaminated by $MgO$ reproducibly increases after the treatment with alkali metals. Taking into account these results, the increasing of $T_c$ of the $MgO$-contaminated $MgB_2$ samples [7] was attributed to metal/metal oxide interfaces, which were formed by alkali metal chemical reduction of $MgO$ inclusions in the samples [7].

This assumption has been supported by observation of superconductivity at 20–34 K in the $Na/NaO_x$ interfaces formed on the surface of bulk sodium core [8]. Later on, a strong indication of the interfacial superconductivity at 20–90 K, 45 K, 100–125 K and 39–54 K has been observed, respectively, in the $Cu/CuO_x$ [9], $Al/Al_2O_3$ [10], $Fe/FeO_x$ [11] and $Mg/MgO$ [12] interfaces.

In this communication, we report on the superconductivity at $\approx 50$ K in the $Mg/MgO$ interface observed by comparative measurements of the ac magnetic susceptibility and dynamic electrical resistance measurements. The interface was prepared by surface oxidation of metallic magnesium under special conditions [12].

## 2. Material and methods

A nearly ball-shaped sample of 99.99%-pure magnesium with a diameter $d \approx 3$ mm, was placed into a quartz pipe with an inner diameter of 4 mm and a wall thickness of 1 mm, and exposed to 20 h oxygen blasting (10–20 ml/min) at $820 < T < 840$ K, which resulted in coating of the sample's surface with magnesium oxide MgO [12]. The pipe was then removed from the furnace, evacuated within 3–5 min to a residual pressure of ~ 1 Pa, sealed, and quenched to 77 K by liquid nitrogen. The prepared Mg/MgO samples sealed in the evacuated quartz ampoules, were stored in liquid nitrogen to prevent their degradation between the measurements, because under normal conditions they are instable [12].

The samples were studied by measuring the complex alternating current (ac) magnetic susceptibility $\chi = \chi' - i\chi''$ and the dynamic electrical resistance. The susceptibility was measured using a low-temperature ac susceptometer involving a mutual inductance technique [13] operating at ac frequency $\nu = 1550$ Hz, ac excitation field $H_{ac} = 0.2$ Oe and temperatures $4.5 \leq T \leq 70$ K.

To avoid sample degradation, the measurements of the Mg/MgO sample were done without unsealing the ampoule. The ampoule with the Mg/MgO sample was mounted on the measurement insert in the ambience of liquid nitrogen. Avoiding the sample warming, the insert was then rapidly dipped into a liquid helium dewar and cooled down to 4.5 K, followed by $\chi(T)$ measurements upon heating at the rate of 1-1.5 K/min.

The ac susceptibility responses to the superconducting transitions of lead and niobium samples prepared in the shape of the samples in study and measured under the same conditions were used for calibration of the susceptometer. The ac susceptibility response to the superconducting transition of an $MgB_2$ sample sealed in an evacuated quartz ampoule was found at 39 K confirming the correct thermometry of the measurement setup.

Dynamic electrical resistance measurements, $dU/dI$, were performed by low frequency (13 Hz) ac lock-in techniques with excitation current amplitude $I_{ac} = 24$ mA, combined with bias dc current $0 \leq I_{dc} \leq 400$ mA. A spring-clamp four-probe method was applied for the measurements. The voltage clamps were stainless steel pins fed through insulated coaxial holes in 6 mm-diameter copper disk-shaped current clamps. The electrical resistance of the current and voltage contacts was lower than 0.01 Ohm. The electric current Joule heating of the sample was not observed during the measurements.

To prevent degradation of the Mg/MgO sample when being mounted, the ampoule with the sample was placed into a foam plastic box filled with liquid nitrogen, where the sample was retrieved from the ampoule and spring-clamped to the electric leads in the measurement insert, followed by dipping the insert into the liquid helium dewar for the $dU(T)/dI$ measurements. In order to avoid the sample degradation, it was never let to be warm above 77 K.

## 3. Results and discussion

Curves 1 in Fig. 1 showing the temperature dependences of $4\pi\chi'(T)$ and $4\pi\chi''(T)$ for the initial magnesium sample, are typical for normal nonmagnetic metals. Namely, the penetration depth $\delta$ of the ac magnetic field at 1550 Hz in magnesium at room temperature is $\approx 2.6$ mm, which is bigger than the sample radius $r \approx 1.5$ mm. As the temperature decreases, the electric conductivity of the sample increases monotonically that leads to a decrease in $\delta$, thus to a monotonic decrease in $\chi'$ [14]. At low temperatures, the electric conductivity of the sample becomes more flat in temperature, and $\chi'(T)$ sets constant (see curve 1 in Fig. 1(a)).

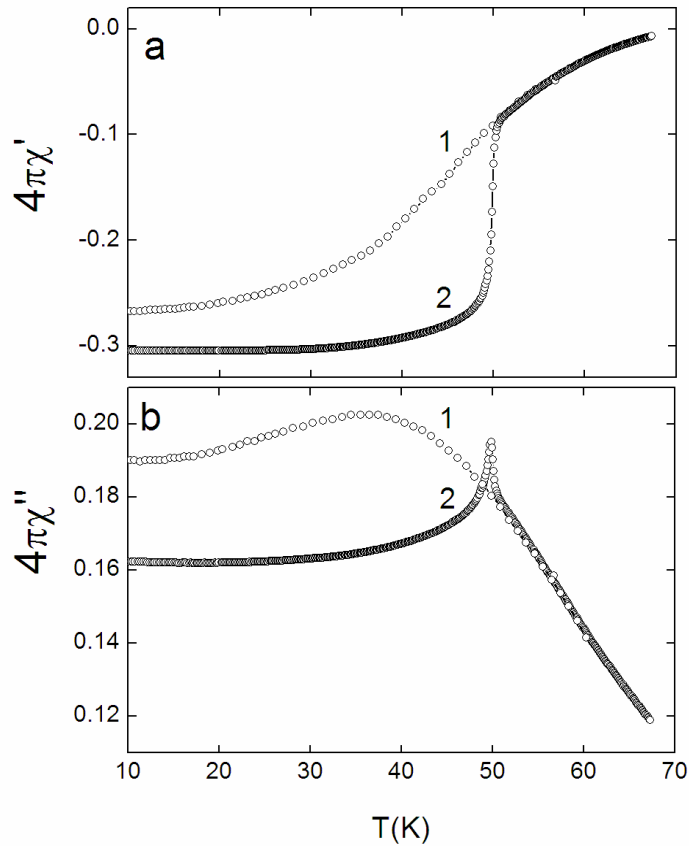

Figure 1. The real part, $4\pi\chi'(T)$ (a) and the imaginary part, $4\pi\chi''(T)$ (b), of the complex ac susceptibility of the initial magnesium sample (Curves 1) and Mg/MgO sample (Curves 2). The measurements were performed in $H_{ac} = 0.2$ Oe at frequency $\nu = 1550$ Hz.

$\chi''(T)$ characterizes the energy dissipation of the ac magnetic field in the sample. This dissipation is also a function of the penetration depth $\delta$ of the ac magnetic field into the sample. The energy dissipation is low at $\delta \ll r$ (high-conductivity metal) and $\delta \gg r$ (insulator). In the intermediate region $\chi''(T)$ has a maximum at $\delta \sim r$ [14]. This maximum is seen on curve 1 in Fig. 1(b) at $T \approx 37$ K.

Curves 2 in Fig. 1 show temperature dependences of $\chi'(T)$ and $\chi''(T)$ for the Mg/MgO sample after applying the oxidation process. As the temperature decreases, $\chi'(T)$ sharply drops at $T \approx 50$ K, accompanied by a sharp maximum in $\chi''(T)$. Since in nonmagnetic materials $\chi'(T)$ and $\chi''(T)$ at fixed frequency are only determined by the temperature dependence of the sample's electric conductivity, the observed drop in $\chi'(T)$ denotes that near the sample surface appears a layer with conductivity much higher than that of Mg.

The observed anomalies in the $\chi'(T)$ and $\chi''(T)$ dependences, curves 2 in Fig. 1, are typical for superconducting materials [15,16], assuming a superconducting transition in the Mg/MgO sample with the onset at $T_c \approx 50$K. It has been shown in [12] that these anomalies are suppressed by the magnetic field, which supports this assumption. Since in the studied temperature range neither metallic Mg nor MgO are superconductors in the bulk, the superconductivity observed in the Mg/MgO sample should be attributed to the Mg/MgO interface formed between the metallic magnesium core and the blanket of MgO at the sample's surface [12].

Fig. 2 summarizes the results of the dynamic electrical resistance measurements. Initially, the voltage terminals were contacted to the internal magnesium core of the Mg/MgO sample by puncturing the oxidized surface layer. The measured $dU(T)/dI$ dependence, curve 1 in Fig. 2, decreases monotonically with cooling regardless of the dc bias current $0 \leq I_{dc} \leq 400$ mA, thus demonstrating a metallic type behavior.

Next, the sample was retrieved from the cryostat into the ambience of liquid nitrogen, and the voltage terminals were reconnected to the oxidized surface of the sample. $dU(T)/dI$ dependencies, measured at the dc bias current values $I_{dc} = 0$, 200, 300 and 400 mA are shown, respectively, by curves 2–5 in Fig.2.

With decreasing temperature, a step-like drop in $dU(T)/dI$ is observed at $T_c \approx 53$ K for $I_{dc} = 0$ (curve 2 in Fig.2). Increasing $I_{dc}$ results in lower $T_c$ and smaller size of the step-like feature, curves 3–5 in Fig.2. We attribute this phenomenon to the superconducting transition in the Mg/MgO sample. Since the phenomenon is not present in $dU(T)/dI$ curves measured on the metal core of the sample (curve 1 in Fig.2), we believe that superconductivity arises in the surface layer of the Mg/MgO sample.

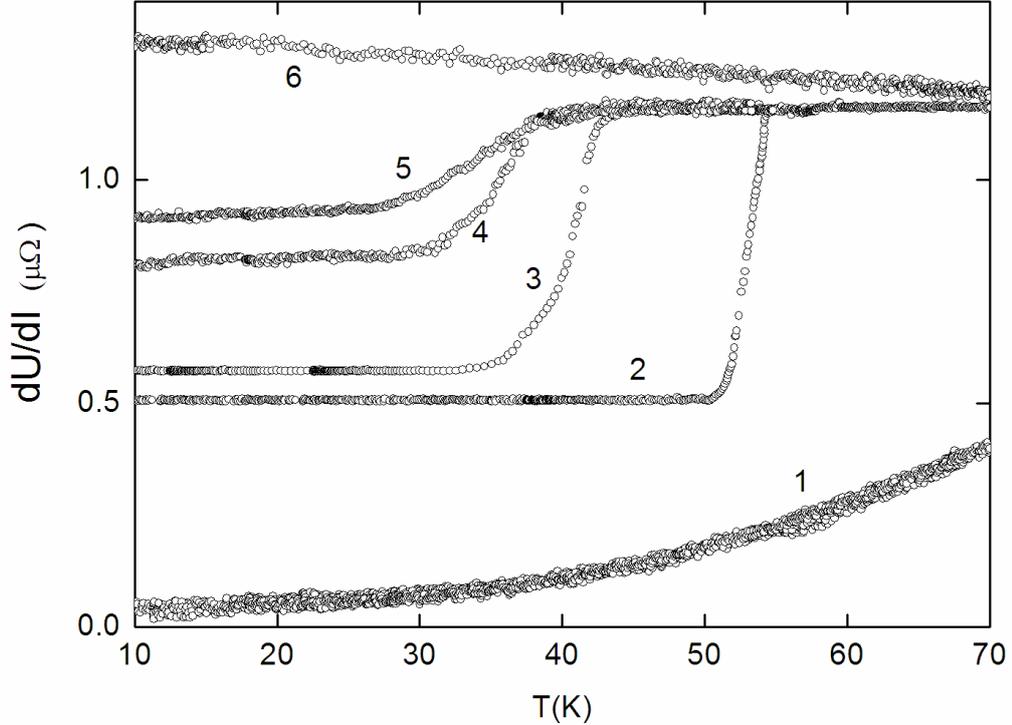

Figure 2. Dynamic electrical resistance, $dU(T)/dI$, of the Mg/MgO sample. Curve 1: Voltage terminals fed through the oxidized sample surface to the internal metallic magnesium core. Curves 2-5: Voltage terminals contacted to the oxidized sample surface; $dU(T)/dI$ measured with the dc bias current $I_{dc}$ = 0 ( Curve 2), $I_{dc}$ = 200 mA ( Curve 3), $I_{dc}$ = 300 mA (Curve 4), $I_{dc}$ = 400 mA (Curve 5). Curve 6: After ~ 30 minutes exposure of the sample to the normal conditions; $I_{dc}$ = 0. In all the measurements the driving ac current amplitude was $I_{ac}$ = 24 mA at frequency $\nu$ = 13 Hz.

As one may see in curves 2, Fig. 1, at temperatures well below $T_c$, $4\pi\chi' > -1$ and $\chi'' > 0$ which indicates an imperfect diamagnetic state, since the ideal case assumes $4\pi\chi' = -1$ and $\chi'' = 0$. This signifies that the superconducting interface does not form a closed surface on the Mg/MgO sample capable of trapping the magnetic flux and keeping it constant, but rather consists of contiguous superconducting regions linked by either normal bridges or Josephson junctions. This is the reason why measurements of the static magnetic moment using a dc SQUID magnetometer[1] did not show a superconducting response for this Mg/MgO sample, indicating no distinguishable fraction of the sample with continuous superconducting surface capable of non-decaying shielding currents. This speculation is supported by the $dU(T)/dI$ measurements which do not show $dU(T)/dI = 0$ at temperatures well below $T_c$, see curves 2–5 in Fig. 2.

---

[1] The ac susceptibility measurements have shown that the superconducting properties of the sample sealed in the evacuated ampoule survive ~15 minutes exposure to room temperature, which enables standard loading routine of the SQUID magnetometer.

Finally, the $dU(T)/dI$ measurement insert with the Mg/MgO sample was removed from the cryostat to the normal atmosphere ambience and kept for ~ 30 minutes at room temperature, followed by $dU(T)/dI$ measurements presented in Fig.2 by curve 6. In contrast to curves 2–5, curve 6 in Fig.2 is monotonic with no sign of the step-like feature which denotes degradation of the superconducting fraction in the Mg/MgO sample due to its instability under the normal conditions. The reason for such instability is apparently the oxygen/magnesium ionic diffusion processes activated under the normal conditions in the Mg/MgO interface, destroying the superconducting fraction. This conclusion is also supported by subsequent susceptibility measurements performed on the degraded Mg/MgO sample, which have given $\chi(T)$ dependences very similar to those of the initial magnesium sample shown by curves 1 in Fig. 1.

The origin of the observed superconductivity is not clear. Unlike the $\chi(T)$ and $dU(T)/dI$ dependences, which change dramatically after the exposure of the sample to the normal conditions due to the decay of the superconducting fraction, no change in the X-ray diffraction patterns of the Mg/MgO sample has been detected [12]. The fraction of the superconducting phase is therefore too small to be detected by the X-ray diffraction technique.

As a possibility, the superconductivity may occur within an interfacial layer separating Mg and MgO phases consisting of unknown nonequilibrium superconducting oxide phase(s), instable under the normal conditions. Alternatively, nanometer-sized metallic clusters with narrow, partially filled energy band at the Fermi level may arise in the metal-oxide interfacial layer. As a result, under specific conditions superconducting pairing in such objects is expected to become very strong, forming a new hypothetical family of high-$T_c$ (~ 150–1000 K) superconductors [17].

We have discovered this phenomenon also in other similar objects based on metals of various groups (Na/NaO$_x$, Cu/CuO$_x$, Al/Al$_2$O$_3$ and Fe/FeO$_x$) [8-11]. This indicates the generality of the observed phenomenon and allows us to conclude that, under certain oxidation and heat treatment conditions, a superconducting layer with relatively high superconducting transition temperature can be formed in metal/metal oxide interfaces for many metals.

4. **Conclusion**

Superconductivity at ≈ 50 K in the Mg/MgO interface has been observed in the ac magnetic susceptibility and dynamic electrical resistance measurements. The interface was formed by the process of surface oxidation of the metallic magnesium under special conditions. These measurements have also shown instability of the superconducting Mg/MgO interface under the normal conditions.

**Acknowledgement**

The work has been supported by RAS Presidium Programs "Quantum physics of condensed matter" and "Thermal physics and mechanics of extreme energy impacts and physics of strongly compressed matter".

**References**

1. N. Reyren *et al.*, Science 317 (2007) 1196-1199.
2. P. Zubko, S. Gariglio, M. Gabay, P. Ghosez, J.-Ma Triscone, Ann. Rev. Cond. Matt. Phys. 2 (2011) 141-165.
3. S. Gariglio, J.-M. Triscone, C. R. Physique 12 (2011) 591-599.
4. J. Pereiro, A. Petrovic, C. Panagopoulosa, I. Božović, Physics Express 1 (2011) 208-241.
5. J. G. Bednorz, K. A. Müller, Zeitschrift für Physik B 64 (2) (1986) 189-193.
6. A. Gozar *et al.*, Nature 455 (2008) 782-785.
7. A.V. Palnichenko, O.M. Vyaselev, N.S. Sidorov, JETP Lett. 86 (2007) 272-274.
8. N.S. Sidorov, A.V. Palnichenko, S.S. Khasanov, Solid State Commun. 150 (2010) 1483-1485.
9. N.S. Sidorov, A.V. Palnichenko, S.S. Khasanov, Phys. C 471 (2011) 247-249.
10. N.S. Sidorov, A.V. Palnichenko, S.S. Khasanov, JETP Lett. 94 (2011) 134-136.
11. N.S. Sidorov, A.V. Palnichenko, I.I. Zver`kova, Phys. C 471 (2011) 406-408.
12. N.S. Sidorov, A.V. Palnichenko, S.S. Khasanov, Solid State Commun. 152 (2012) 443-445.
13. D.-X. Chen, V. Skumryev, Rev. Sci. Instrum. 81 (2010) 025104(1)-025104(10).
14. L.D. Landau, E.M. Lifshitz, *Course of Theoretical Physics,* Vol. 8: *Electrodynamics of Continuous Media* (Nauka, Moscow, 1982; Pergamon, New York, 1984).
15. R.A. Hein, Phys. Rev.B 33 (1986) 7539-7549.
16. T. Ishida, R.B. Goldfarb, Phys. Rev.B 41 (1990) 8937-8948.
17. V.Z. Kresin, Yu.N. Ovchinnikov, Usp. Fiz. Nauk 178 (2008) 449-458.